\input amssym.def 	
\input amssym.tex 	

\documentstyle[a4,11pt,epsf]{article}
\epsfverbosetrue
\newcommand{\figurebox}[2]{\fbox{\vbox to#2in{\hbox to #1in{\hfil} \vfil}}}
\newcommand{\beq}{\begin{equation}}
\newcommand{\eeq}{\end{equation}}

\begin{document}

\begin{titlepage}

\begin{flushright}
Liverpool Preprint: LTH 303\\
Wuppertal Preprint: WUB 93--10\\
hep-lat/9304012\\
21/4/93
\end{flushright}
\vspace{5mm}
\begin{center}
{\Huge A comprehensive lattice study of SU(3) glueballs}\\[15mm]
{\large\it UKQCD Collaboration}\\[3mm]

{\bf G.S.~Bali, K.~Schilling}\\
Physics Department,
Bergische Universit\"at, Gesamthochschule Wuppertal,\\
Gauss Strasse 20,
5600 Wuppertal 1, Germany

{\bf A.~Hulsebos, A.C.~Irving,
C.~Michael, P.W.~Stephenson}\\
DAMTP, University of Liverpool, Liverpool L69~3BX, UK

\end{center}
\vspace{5mm}
\begin{abstract}

We present a study of the $SU(3)$ glueball spectrum
for all $J^{PC}$ values at lattice spacings
down to $a^{-1}=3.73 (6)$ GeV ($\beta=6.4$) using
lattices of size up to $32^4$.
We extend previous studies and show that the continuum limit has
effectively been reached.
The number of clearly identified $J^{PC}$ states has been
substantially increased.
There are no clear signals for spin-exotic glueballs below 3 GeV.
A comparison with current experimental glueball candidates is made.

\end{abstract}

\end{titlepage}

\paragraph{Introduction}\label{introduction}
The extraction of reliable predictions for the glueball spectrum of QCD
remains an important challenge for lattice gauge theory.
As part of a recent programme to study non-perturbative pure SU(3)
gauge theory closer to the continuum
limit, we have obtained new data for glueball masses which confirm that
results of relevance to continuum physics are indeed being achieved with
currently accessible lattices. The low-lying spectrum for pure glue
states below 3 GeV or so has become well established now and awaits progress
with understanding the effects of light-quark mixing.

Glueballs are not particularly light --- they start around 1500
MeV --- and have no non-trivial flavour content. The extraction of a
signal in the presence of vacuum fluctuations
is therefore more difficult than for many other hadrons or for
potentials. In this situation it is highly desirable to perform
coherent measurements over a suitable $\beta$-range, in order not to
be lost in possible systematic effects. In this spirit
we apply here the techniques used by and, in some cases,
pioneered by Michael and Teper (MT)~\cite{MT1,MT2,MT3} and extend
their analysis. They used lattices ranging from $10^4$
to $20^4$, at $\beta$ values
up to $6.2$. In the meantime, there has been progress both in the
available computing power and in the efficiency of
updating algorithms. In this work we have used a
hybrid~\cite{BS1,BS2} of heat-bath and over-relaxation.
The code was specifically developed for the Connection Machine and
was run on an 8K machine at Wuppertal and a 16K machine at Edinburgh.
The key parts of the code, including the group theory, were thus
independent of previous work.
We have concentrated on $\beta=6.4$ on $32^4$ ---  slightly
larger in physical size than the
largest size used by MT, but have also taken data at $\beta=6.0$ and $6.2$
where a direct comparison could be made.

\paragraph{Measurement procedures}\label{measure}

The $\beta=6.4$ results presented in this letter were based on
the measurement of 3220
configurations, each separated by ten sweeps. Every fifth sweep was a
heat-bath
step, the remainder being
Creutz over-relaxation. The data was
obtained in two parts, from hot and cold starts  with at
least 2000 sweeps used to equilibrate in each case. During the
subsequent analysis, described below, a careful check was made that
no residual equilibration effects were present, that both samples were
consistent, and the measurement sampling rate was reasonable
compared with the autocorrelation times.
A direct measurement of the autocorrelation time
gave $\tau\lesssim 20$ sweeps for the correlators of interest.
In order to allow
greater flexibility in analysis and to allow further cross-checking,
the in-line measurements were done in a
rather general way.
They were made on the $(L/2)^3 L$ lattice configuration
obtained from an $L^4$ configuration
by Teper fuzzing ~\cite{Tfuzz} with the link/staple mixing parameter
$\alpha=1.0$.
On each time slice, operator momentum transforms for a variety of oriented
\lq shapes\rq{} and for all cubic orientations were stored for
this level of fuzzing, and for each subsequent
level up to the maximum physically reasonable.
The shapes were as noted in table~\ref{tab:ops}
(see ref.~\cite{MT3} for details and a diagram).
\begin{table}[h]
\centering
\begin{tabular}{clcc}
	{\bf Links}	&{\bf Shapes}	& {\bf Orientations}
	&{\bf Fuzz levels} \\ \hline
	4	& plaquette	&3 	& 1,2,3,4 \\
	6	& rectangle	&6	& 1,2,3	\\
	6 	& chair		&12	& 1,2,3,4 \\
	8	& hand		&48	& 1,2,3	\\
	8	& butterfly	&24	& 1,2,3	\\
\end{tabular}
\caption{\it Glueball operators used. See ref. [3] for a
diagram}
\label{tab:ops}
\end{table}
\noindent Non-zero momentum operators
were used only for the plaquette shape
for which we considered $k_{\mu}=0,\pm 1$ where
$p_{\mu}={{2\pi}\over{L}}k_{\mu}$.
Because of the
initial fuzzing step before measurement,
the space points summed over were spaced by 2 units so that the
momentum eigenvalues $k_{\mu}$ were unique only up to modulo $L/2$.
For small momenta the contamination is expected to decay very fast in
Euclidean time. By studying similar sized lattices in $SU(2)$ we did
confirm this effect but found that the $t=0$ correlations, and hence
$1/0$ ratios showed significant contamination from the high momentum
piece. Because of the variational nature of the calculation (see below)
this did not affect our spectrum results at all.
The total time used to update
the gauge configuration and make these primary measurements was of
order 300 hours (16K CM-200 equivalent). The gauge update time was
$3.8\,\mu$sec per link.

The operator sums for each time-slice were analysed off-line. We studied
Euclidean time correlators for all representations of $O_h$: $A_1$,
$A_2$, $E$, $T_1$ and $T_2$  for both values of parity and C-parity \cite{BB}.
The
relevant projection table is given, for example, in ref.~\cite{CMAPP}.
In addition, we studied Polyakov line correlators (the torelon) as a cross
check on the
string tension. Further data acquired for smeared Wilson loops and for the
topological susceptibility will be presented elsewhere \cite{BS3}. In
the off-line analysis of correlations, a variational approach was used
\cite{MT3} in which a matrix of correlators is formed using, as basis,
the different
relevant operator shapes and fuzzing levels. By diagonalising
the transfer matrix and studying ratios of eigenvalues at consecutive
Euclidean times, one obtains upper bounds on the effective mass (or
energy for non-zero momentum) of the ground state in each channel.
In principle, estimates of excited states can also be made.

There are two cross-checks on the
reliability of the mass values so obtained. First, the overlaps for the
various operators are obtained. For a stable determination, one would
prefer large \lq wave-function\rq{} components carrying the same sign
rather than a delicate cancellation (as a result of a poor choice of basis).
Indeed, we have checked the stability of our results to using smaller
and differing samples of basis operators. The $A_1^{++}$
receives contributions from a  broad range of shapes and fuzzing
levels, while the remaining states receive dominant contributions
from the maximum fuzzing level, mostly from the \lq
hand\rq{} shaped loops (see table ~\ref{tab:ops}).
Second, one expects the ground state in each channel to dominate at large
Euclidean time. We have monitored the {\em difference} between
successive effective masses to find at which $t$ value this becomes
statistically
insignificant. For the determination of errors, we have always used the
bootstrap sampling procedure where the data are organised in bins large
compared with the measured autocorrelation length.\footnote{In fact, for test
observables,  we have used
the bin size dependence of the measured variance to cross-check the direct
measurements of the autocorrelation time quoted above.} For the majority
of states, the effective mass \lq plateau\rq{} identified in
this way starts at time ratio $2/3$.

\paragraph{Results}
Table~\ref{tab:mass} contains the measured effective masses of all
ground state glueballs
which can be studied on a hypercubic lattice.
Where a significant signal was found, the chosen plateau
value and its associated error estimate is indicated by bold
face. Where larger time ratios gave higher masses or where the
plateau was not particularly well established, the error
was conservatively estimated from the next larger time ratio.
In the final column of the table, the glueball masses $m$ are given in
units of the string tension from Wilson loops.
We find that the spectrum
proposed by MT from the average of their large volume
data at $\beta=5.9$, $6.0$ and $6.2$ (~\cite{MT3}, table 7)
is in agreement with our $\beta=6.4$ mass ratios. Our data at $6.0$
(on $16^3\times 32$) and at $6.2$ (on $32^4$) are themselves consistent with
the corresponding results of ~\cite{MT3}. Overall,  we conclude that
the new lattice measurements of the low-lying spectrum at $\beta=6.4$ do
indeed
represent useful information about continuum physics.

\begin{table}[h]
\begin{tabular}{|c||l|l|l|l|l|l|l|}\hline
$O_h$ Rep. & $J^{PC}$\dots &0/1 & 1/2 & 2/3 & 3/4 & 4/5 & $m/\sqrt{\sigma}$ \\
\hline \hline
$A_1^{++}$ &$0^{++}$& 0.604(7) & 0.435(8) & {\bf 0.415(14)}
	& 0.402(20) & 0.38(3) & 3.52(12) \\
$A_2^{++}$ &$3^{++}$& 1.552(15) & 1.06(4) & {\bf 1.05(13)} & 1.0(3)
	& --- &8.9(11) \\
$  E^{++}$ &$2^{++}$& 0.911(5) & 0.653(11) & {\bf 0.620}(17) & 0.61({\bf 3})
	& 0.56(6) & 5.25(25) \\
$T_2^{++}$ &$2^{++}$& 0.914(5) & 0.638(9) & {\bf 0.598}(14) & 0.55({\bf 2})
	& 0.52(4) & 5.07(17) \\
$T_1^{++}$ &$1^{++}$& 1.657(10) & 1.10(3) & {\bf 1.06(8)} & 1.0(2)
	& 0.8(6) & 9.0(7) \\
\hline
$A_1^{-+}$ &$0^{-+}$& 1.155(9) & 0.751(18) & {\bf 0.63}(4) &
0.69{\bf (7)}	& 0.68(14) & 5.3(6) \\
$A_2^{-+}$ &$3^{-+}$*& 2.34(3) & 1.56(18) & 2.6(24) & --- &  & \\
$  E^{-+}$ &$2^{-+}$& 1.265(8) & 0.853(16) & {\bf 0.83(4)} &
0.77(8) &0.9(2) & 7.0(3)\\
$T_2^{-+}$ &$2^{-+}$& 1.284(7) & 0.851(12) & {\bf 0.79}(3) & 0.80({\bf 6})
	& 0.95(19) &  6.7(5)\\
$T_1^{-+}$ &$1^{-+}$*& 1.824(10) & 1.22(3) & 0.99(11) &1.2(5)
&--- &  \\
\hline
$A_1^{+-}$ &$0^{+-}$*& 2.24(2) & 1.31(9) & 0.8(2) & 0.9(6)
	& --- &  \\
$A_2^{+-}$ &$3^{+-}$& 2.68(3) & 1.6(2) & 0.9(5) &---  & & \\
$  E^{+-}$ &$2^{+-}$*& 2.090(18) & 1.23(5) & {\bf 1.2(2)} & 0.5(4) &---
	&10(2) \\
$T_2^{+-}$ &$2^{+-}$*& 1.461(12) & 0.97(2) & {\bf 0.91}(5) & 0.93({\bf 12})
	& 1.0(4) &7.7(10)  \\
$T_1^{+-}$ &$1^{+-}$& 1.188(5) & 0.837(13) & {\bf 0.78}(3) & 0.82({\bf 7})
	& 0.92(14) & 6.6(6)\\
\hline
$A_1^{--}$ &$0^{--}$*& 2.24(3) & 1.55(14) & 1.4(6)
        &--- & &  \\
$A_2^{--}$ &$3^{--}$& 2.70(3) & 1.5(2) & --- & --- & &  \\
$  E^{--}$ &$2^{--}$& 1.715(13) & 1.07(3) & {\bf 1.03(9)} & 1.7(6)
        &--- & 8.7(8) \\
$T_2^{--}$ &$2^{--}$& 1.804(9) & 1.13(3) & {\bf 1.08(9)} & 0.9(3)
	& --- &9.2(8) \\
$T_1^{--}$ &$1^{--}$& 1.845(11) & 1.17(4) & {\bf 1.22(13}) & 0.8(3)
	& 1.2(10) & 9.9(11) \\
\hline
\end{tabular}
\caption{\it Glueball effective masses at $\beta=6.4$, in lattice units.
The last column shows the ratio to the string tension where the quoted
errors arise only from the glueball statistical errors.
 The $J^{PC}$
value displayed labels the lowest
continuum representation that can contribute. Exotic $J^{PC}$ content
is indicated by (*).}

\label{tab:mass}
\end{table}

The restoration of symmetry provides an additional and more stringent
test for continuum physics.
The expected continuum $J^{PC}$ content of the $O_h$
representations can be found,
for example, in ref.~\cite{BB}. In table~\ref{tab:mass}, we indicate
only the lowest possible $J^{PC}$.
We confirm that the $E$ and $T_2$ ground
states (that both contribute to $J=2$ in the
continuum $O(3)$ symmetry group) exhibit the expected degeneracy for
all $PC$ combinations. For $PC=++$ this has been found previously in
ref.~\cite{MT3} for $\beta\geq 6.0$.
A related requirement is the restoration of
the continuum dispersion relation i.e. Lorentz symmetry.
We have been able to test this for
the momenta ${\bf k}^2 = 0,1,2,3$.
A one parameter fit\footnote{For technical reasons,
we have not attempted
a full correlated error analysis.}
of $E_k$
to
\beq
E_{\bf k}a=\sqrt{m_0^2a^2+\left({{2\pi\left|{\bf k}\right|}\over{L}}\right)^2}
\label{ref:drln}
\eeq
yields for the $A_1^{++}$ data $m_0a=.425(12)$ with
$\chi^2/\hbox{\rm dof}=0.55$.
The non-zero momentum results for the mass, though slightly  higher than
the zero momentum value
given in table \ref{tab:mass},
agree well within errors.  These two features give strong support
to our statement that, for the low-lying states, our $\beta=6.4$ results
are effectively measurements of the continuum glueball spectrum.

Different continuum $J^{PC}$ states can contribute to a given $O_h$
representation. {\em A priori} their level ordering is not obvious.
In fact we observe that the $1^{++}$ mass is definitely larger than
the $2^{++}$ mass. By assigning the $1^{++}$ quantum number
to the $T_1^{++}$ lattice state we have assumed the `natural' ordering
$m_{1^{++}}<m_{3^{++}}$. However, the data at $\beta=6.4$ show the
$A_2^{++}$ and $T_1^{++}$ to be approximately degenerate. This
is not the na\"\i ve expectation since the lowest contributing $J^{PC}$
values in
each case are $3^{++}$, $6^{++}$ and $1^{++}$, $3^{++}$, $4^{++}$,
respectively. One possibility is that the $3^{++}$ state is in fact
lower than the $1^{++}$ and so gives a common lightest contribution to
both lattice states. These higher mass states are difficult to observe
cleanly at lower $\beta$ where $ma$ is unhelpfully large.
Indeed this possible degeneracy was not seen by MT at $6.2$ or $6.0$.
Our estimate of $m_{A_2^{++}}/\sqrt{\sigma}$  at 6.4 is consistent
with the data of ref~\cite{MT3} at 6.2, and with both our 6.2 and 6.0 data.
For the $T_1^{++}$ also, we
have a reasonable signal at $6.4$ unlike at $6.2$ where an upper limit only
was obtained ~\cite{MT3}.

\begin{figure}[htbp]
\vspace{16cm}
\includegraphics{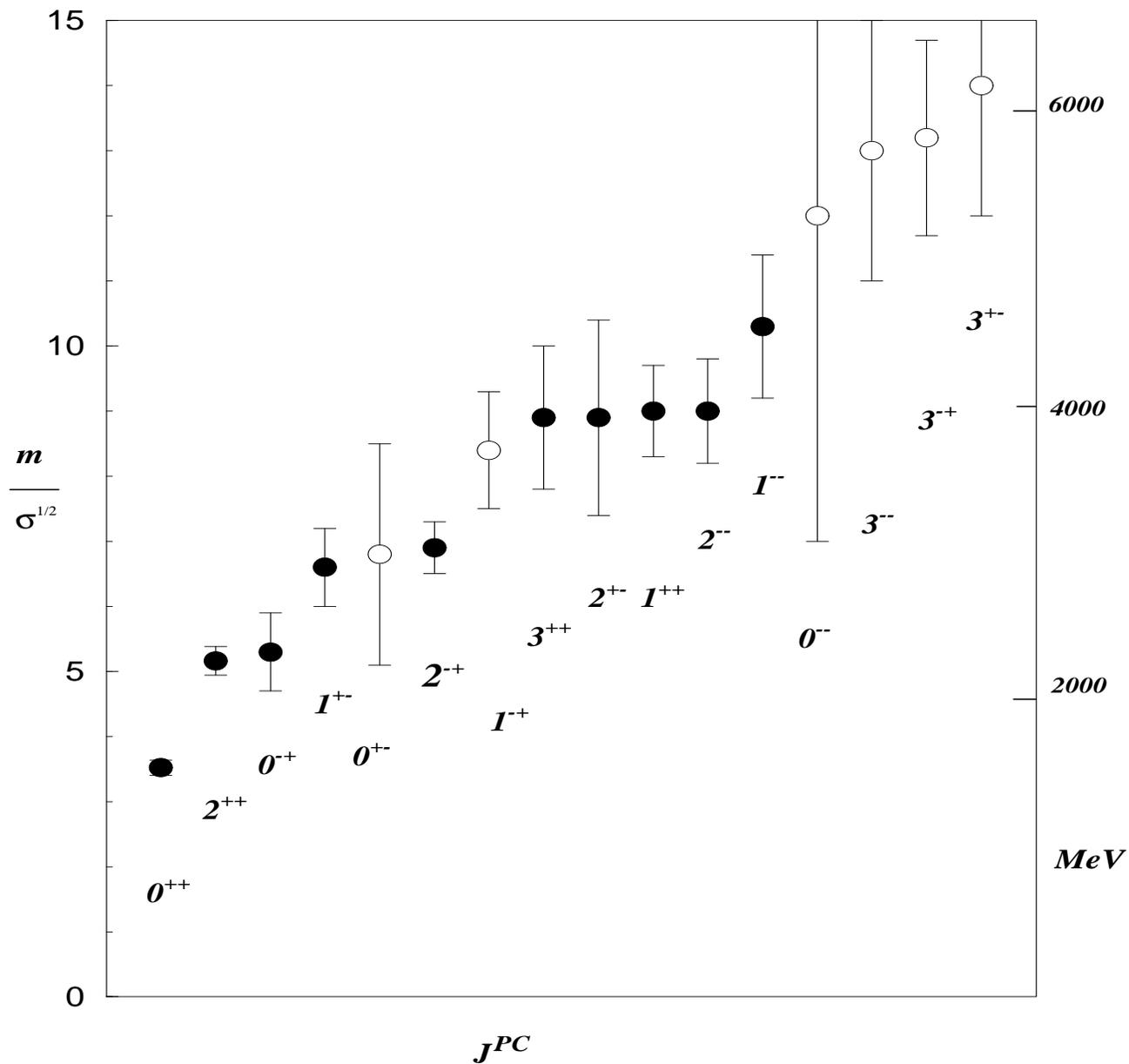}
\caption{The measured glueball spectrum at $\beta=6.4$.
Open symbols represent measured upper limits. The origin of the MeV scale
is described in the text.
}
\vskip 1.0 true cm
\label{fig:gbspect}
\end{figure}

According to table~\ref{tab:mass} we observe reasonably good
signals for 10 states
of different continuum $J^{PC}$ contents.
These are included as the solid circles in fig.~\ref{fig:gbspect}.
Moreover, we determine upper limits for the masses of the remaining
6 states with $J<4$. These are also shown in the figure.
With our lattice resolution ($a^{-1}\approx 3.7$ GeV) and statistics
we are in a position to trace the signals over larger
time separations and achieve
more stringent upper bounds on masses than previously possible.
This improves our capability to separate
low lying glueball states and establish the spectrum order.
The $2^{++}$ is separated by some 6 standard deviations from the
lightest ($0^{++}$) glueball.
Moreover, the $2^{-+}$ glueball is found to be significantly heavier than
the $0^{-+}$.
Above the  $2^{-+}$, five further states have been identified but
their  ordering cannot
yet be determined.

Before proceeding with further interpretation of our results, it is
useful to convert the continuum predictions in the last
column of table~\ref{tab:mass} to  an MeV scale. The extra scale displayed in
fig.~\ref{fig:gbspect} was obtained by
multiplying the latter numbers by a string tension value of $440$~MeV.
Direct comparison with experiment is valid only within the following
two assumptions: (a) that the glueball masses are, for some reason,
insensitive to light quark mixing where this takes place and (b) that
the physical
string tension estimate $(440 \hbox{\rm  MeV})^2$ based on a model for
Regge trajectories is reliable for the pure glue sector.
Some encouragement for this
latter belief is provided by the fact that the scale for light
hadron masses
set in this way is very reasonable.
A recent large scale quenched lattice study of the
light meson and baryon spectrum at similar lattice spacings \cite{GF11}
shows excellent agreement with
experiment, provided sufficiently small
valence quark masses are used for extrapolations.
Using the $\rho$ mass to set the scale at $\beta=6.17$ yields $a^{-1}=2.63(4)$
GeV which is quite consistent with $2.78(5)$~GeV~\cite{UKQCDrc}
and $2.72(3)$~GeV~\cite{BS1} at
$\beta=6.2$ as obtained from the string tension.
Furthermore, values of $\Lambda_{\overline{MS}}$
deduced from independent studies of the $SU(3)$ heavy quark potential
($256\pm 20$ MeV \cite{UKQCDrc} and
$244\pm 8$ MeV \cite{BS2})
are compatible
with the (unquenched) values found in experiment~\cite{Hebbeker}.

To set the scale in  what follows,
we have used string tension values $\sigma a^2$ extracted
from Wilson loops measured on sufficiently large lattices:
$0.168(11)$ at $\beta=5.7$ \cite{Born}.
$0.073(1)$ at $\beta=5.9$ \cite{Born}
$0.0476(7)$ at $\beta=6.0$ \cite{Perant},
$0.0251(8)$ at $\beta=6.2$ \cite{UKQCDrc} and
$0.0138(4)$ at $\beta=6.4$ \cite{BS3}.
It is interesting to note that the
latter value is in good agreement with the
effective string tension deduced from our Polyakov-Wilson line
correlator (torelon)
\beq
\sigma_{\hbox{eff}} a^2=am_{\hbox{tor}}/L= 0.440(20)/32=0.0138(6)\, .
\label{eq:tor}
\eeq
One should remember that the latter is subject to a finite size correction
of order $\pi/3L^2$~ \cite{DFtor} which on our size of
lattice is small ($+0.0010$).

\begin{figure}[htbp]
\vspace{16cm}
\includegraphics{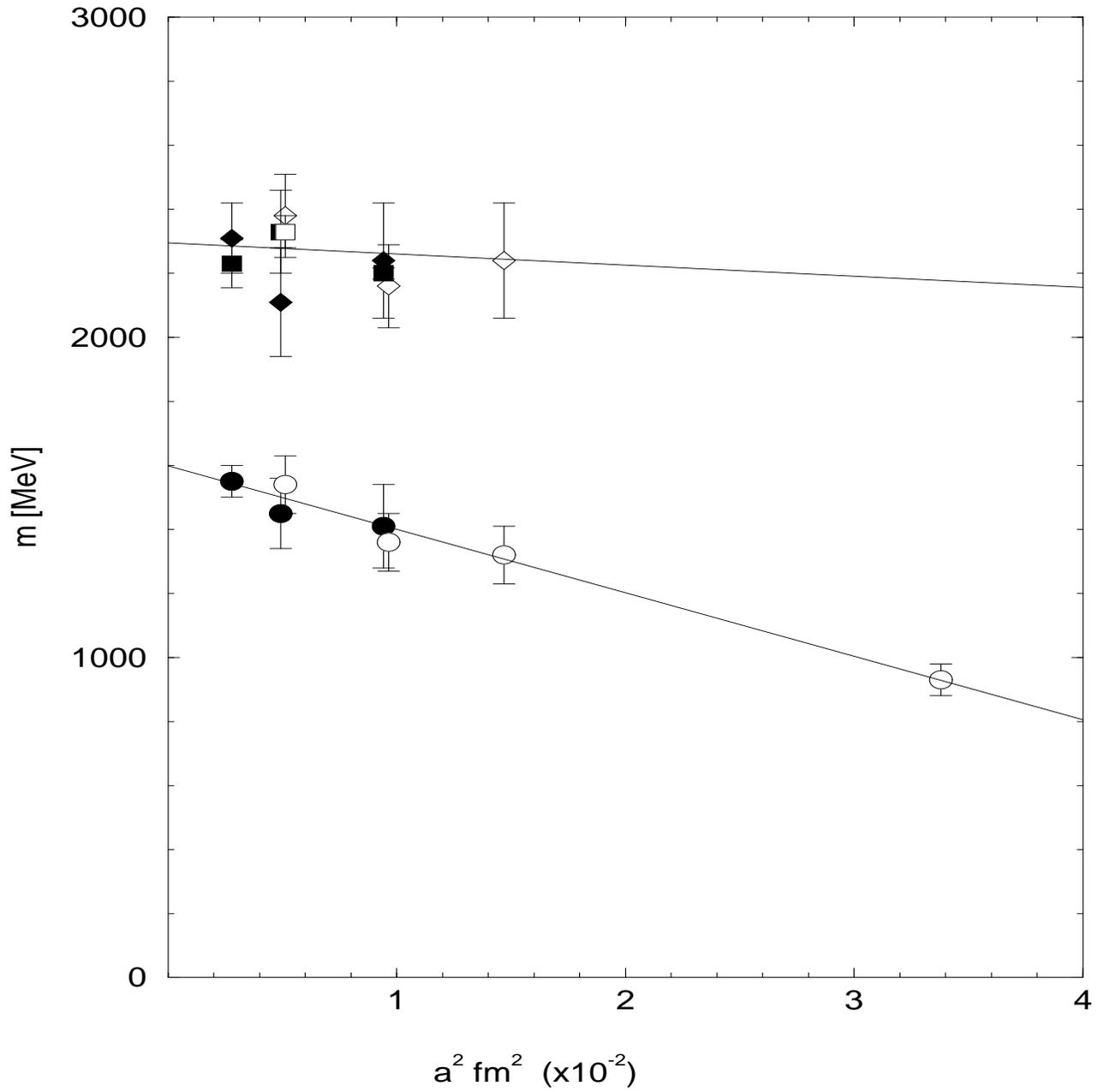}
\caption{Scalar and tensor glueball masses as a function of $a^2$. Full
symbols refer to our data. Open symbols are taken from
refs.~[3,17]. Circles are $A_1^{++}$, diamonds $E^{++}$,
and squares $T_2^{++}$. For clarity data points at the same $\beta$
values have been separated slightly.}
\vskip 1.0 true cm
\label{fig:gbmass}
\end{figure}

In fig.~\ref{fig:gbmass}, we show a new compilation of
scalar and tensor glueball masses, measured at various couplings.
Since the lattice corrections to $ma/\sqrt{\sigma}a$ are expected to be of
order
$a^2$, we display the physical masses
as a function of $a^2$ where physical units on both axes have been
set as described above. The present data is displayed with full symbols
and previous data with open symbols ($\beta=6.2, 6.0, 5.9$ \cite{MT3},
$\beta=5.7$ \cite{DForc}).
We are aware that the string tension results have been obtained by slightly
different methods  but, for present purposes, the ensuing
uncertainties are
small compared to the statistical errors from the glueball masses.
It is clear that, at least for the
scalar glueball, there appears to be little room for uncertainty in any
reasonable extrapolation to $a^2=0$. A linear fit is shown as an example.
The fitted slope to the scalar glueball data
suggests  a systematic error of less than 5\%{} in extrapolation from
$\beta=6.4$,
which would be of the same order as the statistical error.
Because of this, we henceforth use the results at $\beta=6.4$ as an adequate
approximation to the continuum spectrum.

\paragraph{Phenomenological considerations}
With the above scale ($\sqrt{\sigma} = 440$~MeV), the scalar
glueball mass prediction from our
$\beta=6.4$ data is
\beq
m_{0^{++}}=1550\pm 50\,\hbox{\rm MeV}
\label{eq:a1m}
\eeq
where the error here is purely statistical.
This value is in agreement with previous lattice glueball calculations
(e.g. refs.~\cite{BB,MT1,MT2,MT3,Gupta}).
The status of the $G(1560)$ \cite{GAMMS1} as a $0^{++}$ glueball candidate
has recently been considerably strengthened by its independent
observation
in $\bar{p}p\rightarrow 6\gamma$ \cite{CBarrel} where a strong
coupling to the $\eta\eta$, but not to the $\pi^0\pi^0$ channel,
is found. The total width
is $245\pm 50$ MeV. Clearly, the lattice calculation is quite consistent
with this mass.
However, the lattice state is
not far from the broad $q\bar{q}$ state $f_0(1400)$ observed
predominantly in $\pi\pi$ but which also couples to $\eta\eta$ (and
$\gamma\gamma$). Because of the influence of the $f_0(975)$ and the
$K\bar{K}$ channel to which they both couple, the width (a few hundred
MeV) and indeed the very nature of the $f_0(1400)$ is difficult to
establish.
It seems not unlikely that this pure glue state
will suffer mixing and be part of a complex system
involving the above states. Future lattice studies of light quark
mixing will be very illuminating on this point. Pioneering attempts
to study this \cite{Bitar,kfmou} are hampered by the unphysically large
quark masses currently accessible and the difficulty in acquiring
sufficient statistics.

The above energy scale estimate puts the tensor glueball at
\beq
m_{2^{++}}=2270\pm 100\,\hbox{\rm MeV}\, .
\label{eq:T2m}
\eeq
So far, only one experiment has provided evidence of a $2^{++}$
glueball candidate in
this mass range \cite{lindenb}. A series of three $\phi\phi$ states in the
range 2010 to 2340 MeV with widths of 150 to 300 MeV have been
seen but not yet independently confirmed. The next predicted glueball state,
a pseudoscalar at around the same mass (table \ref{tab:mass}), has no
suitable experimental candidates currently.  The search becomes
increasingly difficult at high masses where many states overlap and
many channels are competing.

As pointed out previously (e.g.~ref.~\cite{MT3}),
the prediction of low-lying exotic states (i.e. non $q\bar{q}$ quark
model states) would have interesting
theoretical and phenomenological consequences. Michael \cite{CMAachen}
has recently reviewed the lattice and experimental evidence for
glueball and hybrid states with these quantum numbers.
Our results (table~\ref{tab:mass}) confirm earlier predictions that no exotic
glueball states are expected below about 3 GeV. On the lattice,
each  $O_h$ representation corresponding to an
exotic $J^{PC}$: $0^{--}$, $0^{+-}$, $1^{-+}$ etc.\ can also
receive contributions from higher, but
non-exotic, $J^{PC}$ so identification is
unlikely to be straightforward in the absence of very precise
data.
The strongest exotic signal we have observed is in the $T_2^{+-}$
channel. There is also some evidence of a signal in the
$E^{+-}$ channel. These could correspond to a $2^{+-}$ exotic glueball
at around $3.9\pm .7$ GeV. However, the lowest non-exotic $J^{PC}$
contributing to the $T_2^{+-}$
would be $3^{+-}$ ($5^{+-}$ for the $E^{+-}$) and so no
strong conclusion may be drawn.
The $A_1^{+-}$ and $T_1^{-+}$ channels also show some sort of signal.
These do not satisfy the above criteria for plateau identification
and so we only quote these as upper limits.
Experimental confirmation
of exotic states in the above  mass range is likely to be very difficult.

In conclusion,  we have demonstrated that at
$\beta=6.4$ we are effectively at the
continuum limit
for the quenched glueball spectrum below 3 Gev.
 To be specific, we
observe clear signals for 10 different continuum states.
Thus the ordering of the underlying spectrum is becoming established.
Further improved studies of lattice glueballs are  both
practicable and desirable. In the near future, machines capable of
sustaining
50 to 100 Gflops on QCD will allow a factor
of $\sqrt{50}$ or so reduction in statistical errors and hence greatly
improved effective mass signals.
The present work represents a reduction in lattice spacing by 25\%{} and
an increase in physical volume by 70\%{} over previous studies.
Our results show that there is no need to use larger lattices
or smaller lattice spacings to probe the mass range which is likely to
be of most experimental relevance i.e. below 3 GeV.
However, higher mass states will require larger statistics and lattice
spacings
such that $ma<1$.
In order to estimate the possible influence of mixing effects due to light
quarks it will be vital to have increased  precision of meson and
glueball masses in quenched QCD. This is almost within our grasp.

\paragraph{Acknowledgements}

We thank M.~Teper for helpful conversations.
This research is supported by the UK Science and Engineering
Research Council under grant GR H01236 and the EC under grant
SC1*-CT91-0642.  We are grateful to the
Edinburgh Parallel Computing Centre for access to the CM-200 which
is supported by the Advisory Board to the Research Councils, and
to the Deutsche Forschungsgemeinschaft for support of the
Wuppertal CM-2 project (grant Schi 257/1-4).

\end{document}